\documentclass[a4paper,onecolumn,prb]{revtex4}
\usepackage{amsmath,amsfonts}
\usepackage{graphicx}

\begin{document}
\title{Dynamics of a stretched nonlinear polymer chain}

\author{M. Febbo}
\author{A. Milchev}
\author{V. Rostiashvili}
\author{T. A. Vilgis}
\author{D. Dimitrov}

\affiliation{Max Planck Institute for Polymer Research,
Ackermannweg 10 D-55128 Mainz, Germany}
\date{\today}

\begin{abstract}
We study the relaxation dynamics of a coarse-grained polymer chain
at different degrees of stretching by both analytical means and
numerical simulations. The macromolecule is modelled as a string
of beads, connected by anharmonic springs, subject to a tensile
force applied at the end monomer of the chain while the other end
is fixed at the origin of coordinates. The impact of bond
non-linearity on the relaxation dynamics of the polymer at
different degrees of stretching is treated analytically within the
Gaussian self-consistent approach (GSC) and then compared to
simulation results derived from two different methods: Monte-Carlo
(MC) and Molecular Dynamics (MD).

At low and medium degrees of chain elongation we find good
agreement between GSC predictions and the Monte-Carlo simulations.
However, for strongly stretched chains the MD method, which takes
into account inertial effects, reveals two important aspects of
the nonlinear interaction between monomers: (i) a coupling and
energy transfer between the damped, oscillatory normal modes of
the chain, and (ii) the appearance of non-vanishing contributions
of a continuum of frequencies around the characteristic modes in
the power spectrum of the normal mode correlation functions.
\end{abstract}
\maketitle

\section{Introduction}
The dynamics of a linear polymer was first studied and solved by
Rouse in his seminal paper \cite{Rouse} on the relaxation behavior
of a phantom (ideal) chain of volumeless beads connected by
harmonic (linear) springs. This model, which reveals the
transcendental features of a linear chain, gives all
characteristic dynamic values for nearly free chains whereby the
assumed linearity of the model provides a good description of the
process under investigation. Since this simplified model fails to
retain the physics when the chain is subjected to strong tensile
forces (large stretching), several approaches have been suggested
which take into account the finite extensibility of the chain.
Most of them are based on modifying the underlying force between
monomers, whose nonlinear nature manifests itself for high
extensions. Among the earlier studies that have been tried to
understand the dynamics of highly stretched polymers one should
mention the work of Pincus \cite{Pincus77} which considered
internal modes of a strongly stretched chain using the blob model
\cite{DeGennes79}. Based on this work, Marciano and Brochart-Wyart
\cite{Marciano95}, studied a single chain stretched by a force $f$
applied at the free end and fixed with the other end at the
origin. They demonstrated that stretched chains can be thought as
Rouse chains of impenetrable blobs where all the blobs are
identical. In the long wave-length limit chain dynamics is
described by renormalized Rouse modes of mode number $p$ with a
dispersion relation for the corresponding relaxation times $\tau_p
\propto p^{-2}$. More recently, the work of  Hatfield and Quake
\cite{Hatfield99} presented theoretical calculations and computer
simulations to estimate the effects of tension and hydrodynamics
on $\tau$, the fundamental relaxation time of the polymer.

Apparently none of these studies, however, seems to have reached a
complete and definite description of the problem, in spite of its
fundamental nature. In view of many experiments and technological
applications, the dynamics of highly stretched chains results
fundamental in studying bond rupture of filled and unfilled
elastomers. So far, there no theory based on molecular models for
the fracture of materials exists. Most of the existing theories
are based on phenomenological ideas, which consider directly the
material parameters
\cite{Busfield99,Gurtin00,Gross01,Gross02,Haupt95,Persson08,Hamed94}.
In this work, the main emphasis is drawn on developing molecular
pictures for breaking chains, unfilled and filled elastomers. It
turns out quickly that the basic problems concerning the behavior
of strongly stretched chains are not treated in such a way, that a
systematic study on the chain breaking mechanism can be carried
out. Therefore we concentrated ourselves basically to study the
dynamics of highly stretched chains. To do so, there are two
different ways to tackle this problem. The most obvious
possibility is to introduce constraints, where the lengths of the
chain segments are fixed at any times during the dynamical
processes. Such a theory requires the introduction of Lagrange
multipliers. The theory can be formulated on an exact way, but
nevertheless the solution needs many approximations, which are not
always easy to control. The corresponding stochastic equations are
highly non-linear and do not allow the complete determination of
the Langrange multipliers \cite{Doi}.

To be more specific, we state the problem which we are going to
discuss below in some more details.
Polymer dynamics is usually studied by the so-called Rouse equation, which
has its origin in the corresponding Edwards Hamiltonian for
Gaussian Chains. In discrete notation Gaussian chains are
described by a Gaussian chain statistics, which is defined by the
Hamiltonian
\begin{equation}\label{eq:EdHamiltonian}
    H= k_B T \frac{3}{2b_0^2} \sum_{i=1}^N {\bf b}_i^2
\end{equation}
where $k_BT$ is the thermal energy, ${\bf b}_i = {\bf r}_i - {\bf
r}_{i-1}$ the individual bond vectors of a chain defined by a
connected set of position vectors $\{ {\bf r}_i \} $ and $b_0^2$
their mean value. The Rouse equation, which is the Langevin
equation for the monomer position vector ${\bf r}_i$ is simply
defined by the equation of motion
\begin{equation}\label{eq:Rouse-equation}
\xi_0 \frac{\partial {\bf R}(s,t)}{\partial t} - \frac{3 k_B
T}{b_0^2}\frac{\partial^2 {\bf R}(s,t)}{\partial s^2}={\bf h}(s,t)
\end{equation}
where ${\bf r}_i$ has been represented by continuous space curves
$\textbf{R}(s,t)$ and $s$ is the arc length variable running from 0
to $L$ along chain contour. In last equation, the first term
corresponds to the frictional force,
 $\xi_0$ being the friction coefficient and $\textbf{h}(s,t)$
is the stochastic force, which is taken as a white
noise. The second term is nothing but the elastic force, and is
easily derived by the gradient of the Hamiltonian $H$. The
drawback of this simple stochastic equation is, that the bond
vectors can be stretched to infinity and thus Gaussian chains
overstretch.

This drawback is normally repaired by the use of rigid
constraints, which fix the bond length strictly to $|{\bf b}_i| =
|{\bf r} _i - {\bf r}_{i-1}| = b_0$, but this yields to all the
problems discussed above. The alternative possibility is the use
of potentials, which do not allow the chain to extend to infinity
at high external forces instead of hard constraints. Although this
corresponds to a very different approach the problems become to
some extend more feasible. As an example we mention the FENE -
potential \cite{Kremer}, which is widely used in computer
simulations, to prevent the overstretching of the chains and
avoiding so unphysical conformations. To do so we have proposed a
modified chain potential according to the FENE - potential, where
the chain segments can only stretch to a maximum value, i.e.,
\begin{equation}\label{eq:FENE-model}
    \beta H=- \frac{3}{2} \sum_{i=1}^N \ln\left(1-\frac{b_i^2}{b_0^2}\right)
\end{equation}
where we have chosen here just for the maximum extension of the
bonds the mean bond size from above (in this eq. this is done for
illustration, for the more specific calculations and simulations
we have chosen more general cases). The distribution function for
the end-to-end vector contains Bessel functions that makes it not
very suitable for analytical computations. Nevertheless in a crude
approximation it can be shown, that its asymptotics is of the form
\begin{equation}\label{eq:end-vector-distribution}
    G({\bf R},N)\propto \exp \left(\frac{3 R^2}{2 N b_0^2} \frac{1}{(1-(R/(Nb_0))^2)^{1/2}}\right)
\end{equation}
which shows that the end vector distribution function $G({\bf
R},N)$ tends to zero when the chain is close to its maximum
extension $R = Nb_0$.

Concerning the dynamics we can also extract some features of the
FENE - type potential. For small elongations, $|{\bf b}_i| << b_0$
the FENE - Hamiltonian yields back the Gaussian chain and so, in
the dynamics the classical Rouse equation. On the other hand it
provides the basis for a non-linear dynamical polymer model, which
can be summarized in the (approximate) non-linear stochastic
equation,
\begin{equation}\label{eq:FENE-Rouse-equation}
\xi_0 \frac{\partial {\bf R}(s,t)}{\partial t} - \frac{3 k_B
T}{b_0^2} \dfrac{\frac{\partial^2 {\bf R}(s,t)}{\partial s^2}} {
1-\frac{1}{b_0^2} \left(\frac{\partial {\bf R}(s,t)}{\partial
s}\right)^2 }={\bf h}(s,t)
\end{equation}
Our intention in the present work is to examine the fundamental
aspect of bond nonlinearity of the chain in the high stretched
limit and its impact on the polymer relaxation dynamics with
analytical methods and simulation techniques. To this end we first
employ the Gaussian self-consistent approach. The method has been
extensively used (see e.g. \cite{Timo}) and appears to provide an
adequate approach to study this type of problems. In addition, we
run Monte-Carlo (MC) simulations to compute and compare the
dynamical behavior with that, predicted by the analytical
approach. It is worth noting that the MC scheme does not take into
account mass in the model. In this sense, we want to know to what
extent this is a good description of the problem. To answer this
question, we also performed Molecular Dynamics (MD) simulations
since at high degrees of chain extension the effects of mass and
inertia may not be neglected; therefore we have to modify the
equation of motion as follows:

\begin{equation}\label{eq:FENE-Rouse-equation-mass}
m_b \frac{\partial^2 {\bf R}(s,t)}{\partial t^2} + \xi_0
\frac{\partial {\bf R}(s,t)}{\partial t} - \frac{3 k_B T}{b_0^2}
\dfrac{\frac{\partial^2 {\bf R}(s,t)}{\partial s^2}} {
1-\frac{1}{b_0^2} \left(\frac{\partial {\bf R}(s,t)}{\partial
s}\right)^2 }={\bf h}(s,t)
\end{equation}

The extension by the inertia term is to first instance not
obvious, though it is intuitive. The dynamics of strongly
stretched chains will always get vibrational modes, whenever
non-linear springs are connecting the beads. In contrast to the
Gaussian chains, which can be stretched to infinity, the springs
take at large stretching forces more and more energy, which will
manifest itself in travelling waves along the longitudinal
direction of the stretched string. Of course this FENE Rouse model
is also only an approximation, since rigid constraints cannot be
described rigorously by any (soft) potential. Nevertheless it
removes the basic non-physical features from the Gaussian chain
model. The chain possesses a finite extensibility and the
distribution function does not contain unphysical (overstretched)
states.

Form this equation we can expect several new features compared to
the Rouse equation.
\begin{enumerate}
    \item We can expect diverging time scales as long as the chain becomes
strongly extended, i.e., close to the maximum deformation, where
only chain fluctuations close to the maximum stretching of the
individual bonds matter. These diverging time scales will be
imposed by the singularity in the stretching force term.
    \item We can expect a ``transition" from relaxational dynamics to a
travelling wave dynamics. At low deformation states the dynamics
is mainly ruled by fluctuations where the FENE nature does not
play a crucial role. Thus at low deformations we readily expect a
Rouse type dynamics. At larger stretching ratios, we have
basically a linear chain (like in one dimensional mono atomic
solids) and ``acoustic phonon modes" can be expected, if the
springs were harmonic.
    \item  Since the springs are highly anharmonic we can expect
non-trivial features for the chain dynamics. At high stretching
ratios, i.e., at large stretching forces, we might have strong
effects from the non-linearity of the potential. At small
deformations the Rouse equation is still a valid and useful
approximation, at large deformations, however, the ``Rouse modes"
will interact with each other, they will exchange energy and
damped solitary wave excitations will appear.
\end{enumerate}

However for the main conclusions the form and power of the
singularity in the modified force term in
eq.(\ref{eq:FENE-Rouse-equation-mass}) is not very important. The
main issue is that a singularity at full extension of the bonds,
i.e, the chain appears. Another aspect is that we have so far
ignored the orientation of the chain upon stretching, which will
simplify the problem slightly since we will split the isotropic
variable $\bf{R}(s)$ into its direction perpendicular and parallel
to the force. It is then physically obvious that both components
will respond differently to the strong stretching force.

In the following we will discuss these ideas in more details. In
the first step we will show how simple linearizations will provide
first ideas to the general problem. There we see how the
nonlinearity changes the character of the dynamics completely,
when the chains are strongly elongated. The simple analytic model
will also provide the differences in the behavior of the two
components $R_{\perp}(s)$ and $R_{||}(s)$.  In the second part of
the paper we will use simulations to go beyond the linear
analytical approach. We shall see that the simulation results
provide important new information on the chain dynamics in the
high stretching regime.

\section{Gaussian self-consistent approach}\label{sec:GSC}

To proceed in a first step with the analytical theory we have
chosen to employ a self consistent Gaussian approach, i.e., a
variational technique, which allows to get a first feeling for the
behavior of such strongly elongated chains. The first step is,
that for such stretched chains with finite extensibility we cannot
argue anymore in looking at the isotropic chain variables
themselves. Therefore the chain vectors need to be separated into
components parallel and perpendicular to the stretching direction.
Despite the mean field, and the linearization character, this will
provide some important insight to the physics of the problem.

The dynamics of a stretched polymer chain at high degree of
stretching is essentially nonlinear. This is due to the necessity
of large forces being applied over a short region of space so as
to prevent the chain from stretching indefinitely. While in the
Rouse (linear) model, the linear springs that provide the
connectivity of the chain fail to meet this condition, it is well
satisfied by the anharmonic springs, described by the frequently
used finite extensible nonlinear elastic (FENE) potential.
However, we treat first the relaxation dynamics of a stretched
polymer chain by solving the nonlinear Langevin equation within
the framework of the Gaussian self-consistent approach (GSC). To
do so we considered a polymer chain with $N$ monomers connected by
FENE springs, with one chain end fixed at the origin of
coordinates and pulled by force $f$ at the opposite chain end. Due
to the external force $f$, there emerges a preferred direction in
space leading to symmetry breaking in the problem, caused by
extending the polymer in direction of the acting force. This
effect splits the dynamics of the chain into two directions
\cite{Hatfield99} with respect to monomers motion, ${\bf R}(s,t)$,
one parallel to the force and the other perpendicular to it,
leading to distinct longitudinal and transverse relaxation times.
For the equation of motion of the statistically averaged position,
$\bar{R}_{\parallel}(s,t) \equiv \langle
R_{\parallel}(s,t)\rangle$, in direction parallel to force, we
have

\begin{equation}\label{eq:firstmoment}
 \xi_0 \frac{\partial \bar{R}_{\parallel}(s,t)}{\partial t} + \left\langle
\frac{\delta V[{\bf R}(s,t) - {\bf R}(s-1,t)]}{\delta
R_{\parallel}(s,t)} \right\rangle + \left\langle\frac{\delta
V[{\bf R}(s,t)-{\bf R}(s+1,t)]}{\delta R_{\parallel}(s,t)}
\right\rangle - f \delta_{sN}=0
\end{equation}
where $V[{\bf R}(s,t)-{\bf R}(s \pm 1,t)]$ refers to the
non-linear potential (FENE in this case) which links the nearest
neighbors along the chain backbone and $\xi_0$ denotes the
friction coefficient of a single monomer. We introduce the
notation, $\bar{R}_{\parallel}(s,t) =\langle R_{z}(s,t) \rangle$.
In direction perpendicular to force, the first moments are
$\bar{R}_x(s,t) = \bar{R}_{y}(s,t) \equiv \bar{R}_{\perp}(s,t) =
0$ due to  axial symmetry. Additionally, we present the equations
of motion for time displaced correlation functions; in the
perpendicular direction $C_{\perp}(s,n,t,t') \equiv \langle
R_{\perp}(s,t) R_{\perp}(n,t')\rangle$ we write
\begin{small}
\begin{equation}\label{eq:eqtcorrperpant}
 \xi_0 \frac{\partial C_{\perp}(s,n,t,t')}{\partial t}+\left\langle
R_{\perp}(n,t') \frac{\delta V[{\bf R}(s,t)-{\bf R}(s-1,t)]+V[{\bf
R}(s,t)-{\bf R}(s+1,t)]}{
\delta R_{\perp}(s,t)} \right\rangle =  \\
\langle R_{\perp}(n,t') h_{\perp}(s,t)\rangle
\end{equation}
\end{small}
and for the parallel counterpart $C_{\parallel}(s,n,t,t') \equiv
\langle R_{\parallel}(s,t) R_{\parallel}(n,t')\rangle$ \\
\begin{small}
\begin{equation}\label{eq:eqtcorrparant}
 \xi_0 \frac{\partial C_{\parallel}(s,n,t,t')}{\partial t}+\left\langle
R_{\parallel}(n,t') \frac{\delta V[{\bf R}(s,t) - {\bf R}(s-1,t)]
+ V[{\bf R}(s,t) - {\bf R}(s+1,t)]}{\delta R_{\parallel}(s,t)}
\right> -f
\bar{R}_{\parallel} (n,t')\delta_{sN}=\\
\langle R_{\parallel}(n,t') h_{\parallel}(s,t) \rangle
\end{equation}
\end{small}
where $h_{\parallel}(s, t)$ and $h_{\perp}(s, t)$ are random
Gaussian $\delta$-correlated forces in parallel and perpendicular
directions respectively. We also study the equal time correlation
functions $C_{\parallel,\perp}(s,n,t,t)=C_{\parallel,
\perp}(s,n,t)$ which in the steady state regime do not depend on
the time $t$. For the perpendicular direction, we have
\begin{eqnarray}\label{eq:eqtcorrperp}
 \xi_0 \frac{\partial
C_{\perp}(s,n,t)}{\partial
t}+K_{\perp}(s,s-1)[C_{\perp}(n,s,t)-C_{\perp}(n,s-1,t)
]+K_{\perp}(s,s+1) [C_{\perp}(n,s,t)-C_{\perp}(n,s+1,t) ] \nonumber \\
+K_{\perp}(n,n-1) [C_{\perp}(s,n,t)-C_{\perp}(s,n-1,t)
]+K_{\perp}(n,n+1) [C_{\perp}(s,n,t)-C_{\perp}(s,n+1,t) ] =2
\delta_{ns} k_B T \label{C_perp}
\end{eqnarray}
and for the parallel,
\begin{eqnarray}\label{eq:eqtcorrpar}
 \xi_0 \frac{\partial
A_{\parallel}(s,n,t)}{\partial t}+K_{\parallel}(s,s-1)
[A_{\parallel}(n,s,t)-A_{\parallel }(n,s-1,t)
]+K_{\parallel}(s,s+1)
[A_{\parallel}(n,s,t)-A_{\parallel}(n,s+1,t) ] \nonumber\\
+K_{\parallel}(n,n-1) [A_{\parallel}(s,n,t)-A_{\parallel}(s,n-1,t)
]+K_{\parallel}(n,n+1)[A_{\parallel}(s,n,t)-A_{\parallel}(s,n+1,t)]
=2 \delta_{ns} k_B T \label{C_parallel}
\end{eqnarray}
Here, we define $A_{\parallel}(s,n,t) = C_{\parallel}(s,n,t) -
\bar{R}_{\parallel}(s,t) \bar{R}_{\parallel}(n,t)$, and the
effective nearest neighbor constant,
$K_{\parallel,\perp}(s,s\pm1)$, is defined by eq.
(\ref{Constant_K}) in the Appendix. The relationship $\langle 2
R_{\perp}(n,t) h_{\perp}(s,t) \rangle = 2 \delta_{ns} k_B T $ is
also derived explicitly in Appendix \ref{ap:2}.

\subsection{First cumulant expansion}\label{sec:Gauss-preav}

Consider first the term $\left \langle \frac{\delta V[{\bf R}(s,t)
- {\bf R}(s-1,t)]}{\delta R_{\parallel}(s,t)} \right\rangle$ which
is the  average force experienced by each monomer. The calculation
yields
\begin{eqnarray*}
\left \langle \frac{\delta V[{\bf R}(s,t) - {\bf
R}(s-1,t)]}{\delta R_{\parallel}(s,t)}\right\rangle  = \left
\langle \frac{\delta}{\delta R_{\parallel}(s,t)} \int d^3 r
\delta[{\bf R}(s,t)-{\bf R}(s-1,t)-r] V(\vec{r})
 \right\rangle =\\
  = - \int d^3 r \left\langle \frac{\partial}{\partial r_{\parallel}}\delta[{\bf R}(s,t) -
{\bf R}(s-1,t)-r]\right\rangle V(\vec{r})
 = \int d^3 r \langle \delta[{\bf R}(s,t) - {\bf R}(s-1,t)-r]\rangle
\frac{\partial V(\vec{r})}{\partial r_{\parallel}}
\end{eqnarray*}

The key point is then to determine the difference $\langle \delta[
{\bf R}(s,t) - {\bf R}(s-1,t)-r] \rangle$. This is accomplished by
employing the following crucial approximation $\langle \delta
[{\bf R}(s,t) - {\bf R}(s-1,t)-r]\rangle \approx \delta [\langle
{\bf R}(s,t)-{\bf R}(s-1,t)-r\rangle ]$ which comes from retaining
only the first term in the cumulant expansion. Taking into account
that $V(\vec{r}) = - \frac{1}{2} k_F b_0^2 \ln(1 -
\frac{\vec{r}^2}{b_0^2})$ and $\vec{r}^2 = r_{\parallel}^2 +
r_{\perp}^2$ one arrives at
\begin{eqnarray*}
 \int d^3 r \langle \delta [{\bf R}(s,t) - {\bf R}(s-1,t)-r]\rangle
\frac{\partial V(\vec{r})}{\partial r_{\parallel}} \approx k_F
\frac{\bar{R}_{\parallel}(s,t)-\bar{R}_{\parallel}(s-1,t)}{1-\left[\frac{\bar{R}
_{ \parallel} (s, t) - \bar{R}_{\parallel}(s-1,t)}{b_0}\right]^2}
\end{eqnarray*}

\subsection{Steady state solution}\label{sec:steady-state}

In order to obtain the steady state solution of
eq.(\ref{eq:firstmoment}), we set $\frac{\partial
\bar{R}_{\parallel}(s,t)}{\partial t}=0$. Then, we can find
$\bar{R}_{\parallel}(1)$ by adding up the $N$ terms for the
$\bar{R}_{\parallel}(s)$, requiring that the boundary conditions
of the problem are $\bar{R}_{\parallel}(0)=0 $ (first monomer is
fixed at the origin of coordinates), and
$\bar{R}_{\parallel}(N+1)-\bar{R}_{\parallel}(N)=0 $ (free end
condition).

\begin{figure}[htb]
\centerline{\includegraphics[scale=0.6]{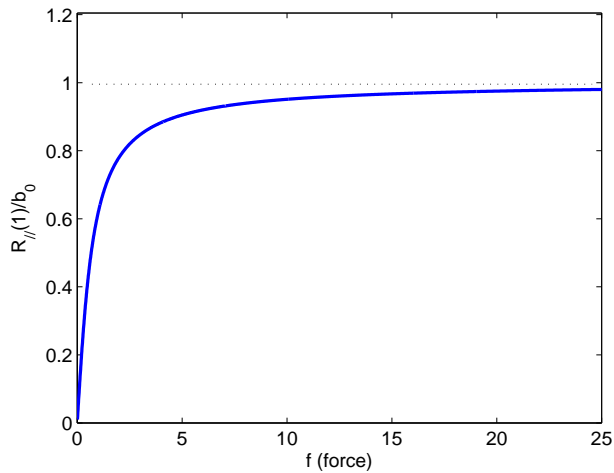}}
\caption{Variation of the bond length between adjacent beads in a
chain with applied tensile force $f$. Eventually, for large forces
the bond length approaches asymptotically its maximum value
$b_0$.}\label{fig:1}
\end{figure}

Eventually we get
\begin{equation}
f=k_{F}\frac{\bar{R}_{\parallel}(1)}{1 -
\left(\frac{\bar{R}_{\parallel}(1)}{b_0} \right)^2}
\end{equation}
from which one can easily derive $\bar{R}_{\parallel}(1) =
\frac{k_F b_0^2}{2 f} \left(-1+\sqrt{1+ \frac{4 f^2}{k_F^2 b_0^2}}
\right)$. In Figure (\ref{fig:1}) we plot the elongation
$\bar{R}_{\parallel}(1)$ as a function of $f$.

Applying an iteration procedure, a general expression for
$\bar{R}_{\parallel}(s)$ can be found, $\bar{R}_{\parallel}(s) = s
\bar{R}_{\parallel}(1),\ 1 < s \leq N$. Setting $\frac{\partial
C_{\parallel,\perp}(s,n,t)}{\partial t}=0$, the steady state
solution for the equal time correlators can also be obtained.
Then,  for the perpendicular direction, the final equation reads
\begin{eqnarray}\label{eq:eqtcorrperpsteady}
 K_{\perp}(s,s-1) [C_{\perp}(n,s) - C_{\perp}(n,s-1)] + K_{\perp}(s,s+1)
[C_{\perp}(n,s) - C_{\perp}(n,s+1)] \nonumber\\
+ K_{\perp}(n,n-1) [C_{\perp}(s,n) - C_{\perp}(s,n-1) ] +
K_{\perp}(n,n+1) [C_{\perp}(s,n) - C_{\perp}(s,n+1)] = 2
\delta_{ns} k_B T,
\end{eqnarray}
and for the parallel
\begin{eqnarray}\label{eq:eqtcorrparsteady}
 K_{\parallel}(s,s-1) [A_{\parallel}(n,s) - A_{\parallel}(n,s-1)] +
K_{\parallel}(s,s+1) [A_{\parallel}(n,s,t) - A_{\parallel}(n,s+1)] \nonumber\\
+ K_{\parallel}(n,n-1) [A_{\parallel}(s,n) - A_{\parallel}(s,n-1)]
+ K_{\parallel}(n,n+1) [A_{\parallel}(s,n) - A_{\parallel}(s,n+1)]
= 2 \delta_{ns} k_B T
\end{eqnarray}

The solution for the time-independent correlator $C_{\perp}(s,n)$
is found by rearranging eq. (\ref{eq:eqtcorrperpsteady}) which
makes it possible to take its continuum limit and rewrite it in a
more convenient way. Eventually, we get
\begin{eqnarray}\label{eq:diffeqetimepar}
 K_{\perp} \left(\frac{\partial^2 C_{\perp}(n,s)}{\partial s^2} +
\frac{\partial^2 C_{\perp}(n,s)}{\partial n^2}\right) = - 2
\delta(n-s) k_B T
\end{eqnarray}
where  $K_{\perp}(s,s\pm1) = K_{\perp}(n,n\pm1) \equiv K_{\perp} =
\frac{k_F}{1 - \left(\frac{\bar{R}_{\parallel}(1)}{b_0} \right)^2
}$. It is better here to define this constant by the relative
degree of stretching $\lambda = \frac{N
\bar{R}_{\parallel}(1)}{\sqrt{N}b_0}$ and its maximum value,
$\lambda_{max}=\frac{N b_0} {\sqrt{N}b_0} = \sqrt{N}$ (where $N b_0$ denotes
the total contour length of the chain).
Thus one can rewrite $K_{\perp} =
\frac{k_F}{1 - \left( \frac {\lambda} {\lambda_{max}} \right)^2}$.

Equation (\ref{eq:diffeqetimepar}) is a linear equation which can
be easily solved for example by means of the Green's function
method \cite{Morse}. In the same way, one can rearrange equation
(\ref{eq:eqtcorrparsteady}) and proceed in a similar fashion. The
final  solutions reads
\begin{equation}\label{eq:solsteadyperp}
C_{\parallel,\perp}(n,s)=\sum_{p=1} \frac{4}{N} \frac{k_B
T}{K_{\parallel,\perp}} \frac{2N^2}{(2p-1)^2 \pi^2} \sin \left(
\frac{(2p-1) \pi n}{2 N}\right)\sin \left( \frac{(2p-1) \pi
s}{2N}\right)
\end{equation}
where $ K_{\parallel}(s,s\pm1) = K_{\parallel}(n,n\pm1) \equiv
K_{\parallel} = k_F\frac{1 +
\left(\frac{\lambda}{\lambda_{max}}\right)^2} {\left[1 -
\left(\frac {\lambda} {\lambda_{max}} \right)^2 \right]^2}$.
These two coupling (effective spring) constants behave therefore differently in the
two directions and depend in a detailed way on the deformation ratio.
Both spring constants show a singularity at the maximum deformation.
We will use these results again in the last section of these paper, when we
discuss the limiting cases and the physics of the numerical results, which will
follow later.

\subsection{Time displaced correlators}

The steady state solution for $C_{\parallel,\perp}(n, s)$ offers
an initial condition from which the time dependent problem can be
solved. Following the GSC method, we are now in a position  to
solve the equations of motion also for $C_{\perp}(s,n,t,t')$.
Using the causality conditions of physical systems $\langle
R_{\perp}(n,t') h_{\perp}(s,t)\rangle = 0$ for $t'<t$, and the
same calculations outlined in the previous section, after taking
the continuum limit, one finally arrives at
\begin{equation}
\xi_0 \frac{\partial C_{\perp}(n,s,t,t')}{\partial t} = K_{\perp}
\left(\frac{\partial^2 C_{\perp}(n,s,t,t')}{\partial s^2}\right).
\end{equation}
This is one-dimensional diffusion equation that can easily be
solved to give
\begin{equation}\label{eq:soldynamicperp}
C_{\perp}(n,s,t,0) = \sum_{p=1} A_p \exp\left ( -\beta_p \frac
{K_{\perp}} {\xi_0} t \right ) \sin \left(\frac{(2p-1) \pi s}{2
N}\right)
\end{equation}
where $\beta_p=\frac{(2p-1)^2 \pi^2}{4 N^2}$ and we have chosen
$t'=0$ without loss of generality.  The amplitudes $A_p$ may be
determined from the initial condition since $C_{\perp}(n,s,0,0) =
C_{\perp}(n,s)$. Equating the expression for $C_{\perp}(n,s,t,0)$
at $t=0$, (eq. \ref{eq:soldynamicperp}), and $C_{\perp}(n,s)$, one
arrives at
\begin{equation}
A_p = \frac{4}{N} \frac{k_B T}{K_{\perp}} \frac
{2N^2}{(2p-1)^2\pi^2} \sin \frac{(2p-1) \pi n}{2 N}.
\end{equation}
Evidently,  $A_p$ is a function of $(p,n)$ since
$C_{\perp}(n,s,t,0)$ is a function of $(n,s,t)$. Now, it is
possible to write down the final expression for
$C_{\perp}(n,s,t,0)$ as
\begin{equation}\label{eq:soldynamicpar2}
 C_{\perp}(n,s,t,0) = \sum_{p=1} \sum_{m=1} \frac{4}{N} \frac{k_B T}{K_{\perp}}
  \frac{\delta_{pm}}{\frac{(2p-1)^2 \pi^2}{2N^2}+\frac{(2m-1)^2 \pi^2}{2N^2}}
e^{-\beta_p \frac{K_{\perp}}{\xi_0} t} \sin \left( \frac{(2p-1)
\pi n}{2 N}\right) \sin \left( \frac{(2m-1) \pi s}{2 N}\right)
\end{equation}
Following the same reasoning one can obtain an identical solution
for the parallel direction too by replacing $K_{\perp} \rightarrow
K_{\parallel}$. From the last equation, the (Rouse) mode
amplitudes can be calculated in a straightforward manner:
\begin{equation}\label{eq:modecorrelperp}
 \langle X_{p\perp,\parallel}(t) X_{m\perp,\parallel}(0)\rangle =
 \frac{k_B T}{K_{\perp,\parallel}} \delta_{pm} \frac{2N}{(2p-1)^2\pi^2}
\exp \left(-\beta_p \frac{K_{\perp,\parallel}}{\xi_0} t\right ) =
  \frac{k_B T}{k_{p\perp,\parallel}} \delta_{pm}
\exp \left( -\frac{t}{\tau_{p\perp,\parallel}}\right),
\end{equation}
 where
$\tau_{p\perp,\parallel} =
\frac{\xi_{p\perp,\parallel}}{k_{p\perp,\parallel}} = \frac{4
N^2\xi_0}{(2p-1)^2 \pi^2 K_{\perp,\parallel}}$, is the
perpendicular (parallel) relaxation time and $k_{p\perp,\parallel}
= \frac{(2p-1)^2\pi^2}{2 N} K_{\perp,\parallel}$ and
$\xi_{p\perp,\parallel} = 2 N \xi_0$.

One should stress that the relaxation time {\em decreases} in both
directions as one increases the degree of stretching since $\tau
\propto 1 - \left ( \frac {\lambda} {\lambda_{max}} \right)^2 $.
This behavior can be understood, if we take into account that the
stretching of the chain makes the springs effectively stiffer.

We must also point out that within the present approach the chain
behaves effectively as Gaussian, $\tau\propto N^2p^{-2}$, which is
plausible since chain stretching rapidly reduces the role of the
excluded volume interactions between the beads. This feature
reveals the nature of the approximation that we have used (see
above) which essentially drives out the coupling between different
modes but still retains the physical intuition of a zero
relaxation time for maximum stretching.

\section{Simulation results}

To test the validity of the approximations made in the previous
section, we performed computer simulations. Two different
numerical schemes have been chosen to test the relaxation dynamics
of stretched polymer chains and also to test the validity of
analytic predictions: Monte-Carlo and Molecular Dynamics. Physical
intuition suggests that for highly stretched polymer chains the
two methods are expected to give different results. This
difference stems from whether the inertia term in the simulations
is neglected or not, which for large extensions of the polymer
plays a decisive role. Since Monte-Carlo simulations do not
consider mass in the model, the corresponding molecular motion of
the chain must be of an overdamped nature (oscillations are not
possible). In contrast, Molecular Dynamics takes inertia into
account and it is possible to pass from an overdamped motion
(small extensions) to an underdamped one (large extensions), in
which oscillations can occur. In what follows we analyze the
simulation results derived by both methods, and compare them to
the predictions of the previous section.

\subsection{Monte-Carlo simulations}

We perform the simulations with an off-lattice coarse-grained
bead-spring model which has been frequently used before for
simulation of polymers \cite{Corsi}. Therefore here we will only
describe briefly its relevant features.   The effective bonded
interaction between nears-neighbor monomers is described by the
FENE (finitely extensible nonlinear elastic) potential.
\begin{equation}
U_{FENE}= -K(1-l_0)^2ln\left[1-\left(\frac{l-l_0}{l_{max}-l_0}
\right)^2 \right] \label{fene}
\end{equation}
with elastic constant $K=20$. The maximum bond extension is
$l_{max}=1$, the mean bond length $l_0 =0.7$, and the closest
distance between neighbors $l_{min} =0.4$

The nonbonded interactions between monomers are described by the
Morse potential.
\begin{equation}
\frac{U_M(r)}{\epsilon _M}
=\exp(-2\alpha(r-r_{min}))-2\exp(-\alpha(r-r_{min}))
\end{equation}
with $\alpha = 24, r_{\mbox{min}}=0.8, \epsilon_M/k_BT=1.0$.

The size of the box is  $64\times 64\times 64$ and the chain
length $N=32$. Of course, a larger number of monomers could in
principle be used but this is not necessary since the most
relevant features of the solution can be well illustrated with
this number of monomers. The standard Metropolis algorithm was
employed to govern the moves with  self avoidance automatically
incorporated in the potentials. In each Monte Carlo update, a
monomer is chosen at random and a random displacement is attempted
with $\Delta x\;$,$\Delta y\;$,$\Delta z\;$ chosen uniformly from
the interval $-0.5\le \Delta x,\;\Delta y,\;\Delta z\le 0.5$. The
transition probability for the attempted move is calculated  from
the change $\Delta U$ of the potential energy $U=U_{FENE} + U_M$
as $W = \exp(-\Delta U/k_BT)$. As usual for the standard
Metropolis algorithm, the attempted move is accepted if $W$
exceeds a random number uniformly distributed in the interval
$[0,1)$. We employ $10^4$ runs of $2^{16}$ Monte-Carlo steps in
each program run.

\subsubsection{MC Simulation results}

During the simulation we compute $\tau_{\parallel}$ and
$\tau_{\perp}$, and the time displaced correlation functions
between different modes, $C_{pq}(t) = \langle
X_{p\parallel,\perp}(t) X_{q\parallel,\perp}(0) \rangle$ (where
$p,q$ label the mode numbers) as a function of the relative degree
of stretching $\lambda/\lambda_{max}$. Here, the average is taken
over different intervals of time $t$.

It is straightforward to compute also theoretically this quantity
in the case of Gaussian chains without excluded volume
interactions \cite{Doi}. This leads us to
\begin{equation}
C_{pq}(t)=\textit{C} \ \delta_{pq}
\exp\left(\frac{-t}{\tau_p}\right) \,p=1..N
\end{equation}
where $C$ is its initial value. For the relaxation time $\tau_p$
we obtain
\begin{equation}
\tau_p = \frac{N^2 \xi_0 b_0}{3 k_B T (p-1/2)^2 \pi^2}
\end{equation}
which clearly shows the $\tau_p \propto N^2 p^{-2}$ dependence of
the relaxation time due to the harmonic nature of the chain
interactions. Obviously, there is no $p=0$ translational mode
since the chain, fixed at both ends, cannot diffuse. For the
highly stretched case it is not clear how the chain will behave,
since the finite extensibility plays a fundamental role and the
forces along the chain are non-linear.

\begin{figure}[htb]
\centerline{\includegraphics[scale=0.6]{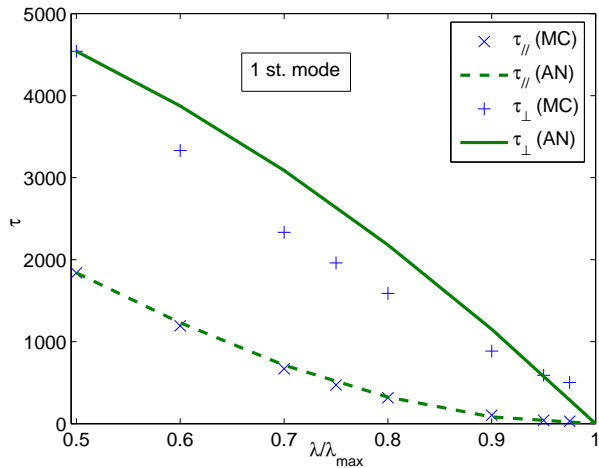}}
\caption{Monte-Carlo simulations of the first {\bf mode}
relaxation time for the perpendicular $\tau_{\perp}$ and parallel
direction $\tau_{\parallel}$ (MC steps) as a function of relative
degree of stretching $0.5<\lambda/\lambda_{max}<1$. Comparison
with analytical (AN) and numerical results
(MC).}\label{fig:MC1time}
\end{figure}

We perform simulations for stretching degrees of $0.5 <
\lambda/\lambda_{max} < 1$ which are shown in Figure
\ref{fig:MC1time}. One can see an almost perfect agreement between
the theoretically predicted (GSC) and simulated (Monte-Carlo)
values of the relaxation time $\tau_{1\parallel}$ of the first
mode in parallel direction. For the perpendicular direction the
matching is not so good but there still exists a qualitative
agreement between analytical and numerical results (the
discrepancy at $\lambda/\lambda_{max} \rightarrow 1$ is due to
numerics since the acceptance rate of the elementary displacements
depends strongly of the attempted jump distance $\Delta x,\Delta
y,\Delta z$).
The relevant feature here is that the relaxation time in both
directions decreases as the chain is gradually stretched to its
final limit. Evidently, the more stretched the chain is, the
``stiffer" it becomes and, consequently, the relaxation time goes
down. There is also a pronounced difference between the parallel
and perpendicular relaxation times for the same $\lambda /
\lambda_{max}$. The correlations in parallel direction decay
faster (shorter relaxation time) than in perpendicular direction.
However, this difference tends to disappear as the degree of
stretching increased.

\begin{figure}[htb]
\centerline{\includegraphics[scale=0.85]{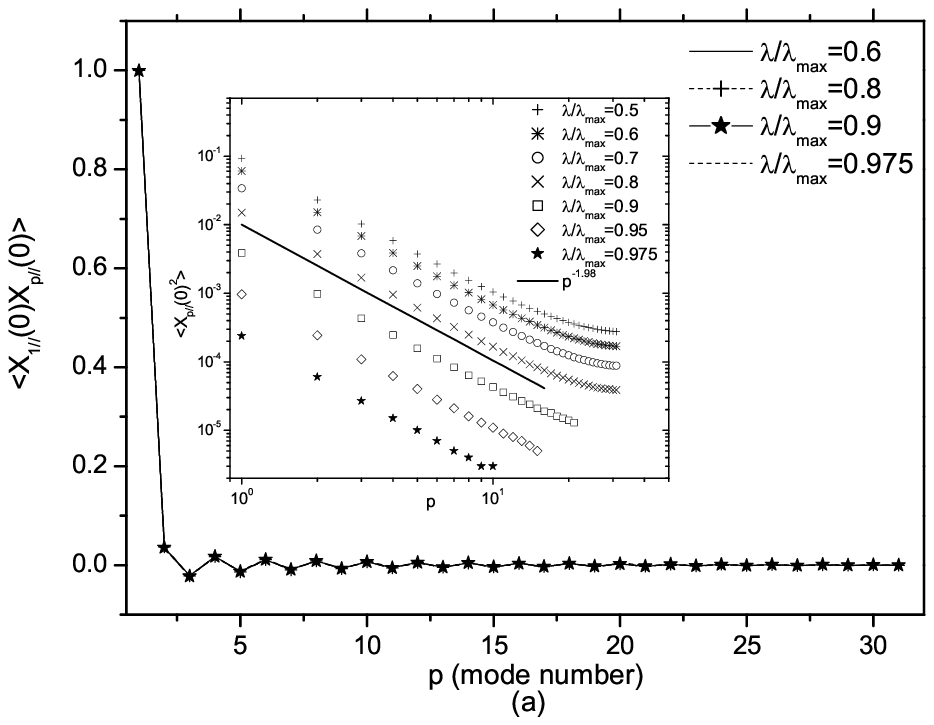}
\includegraphics[scale=0.85]{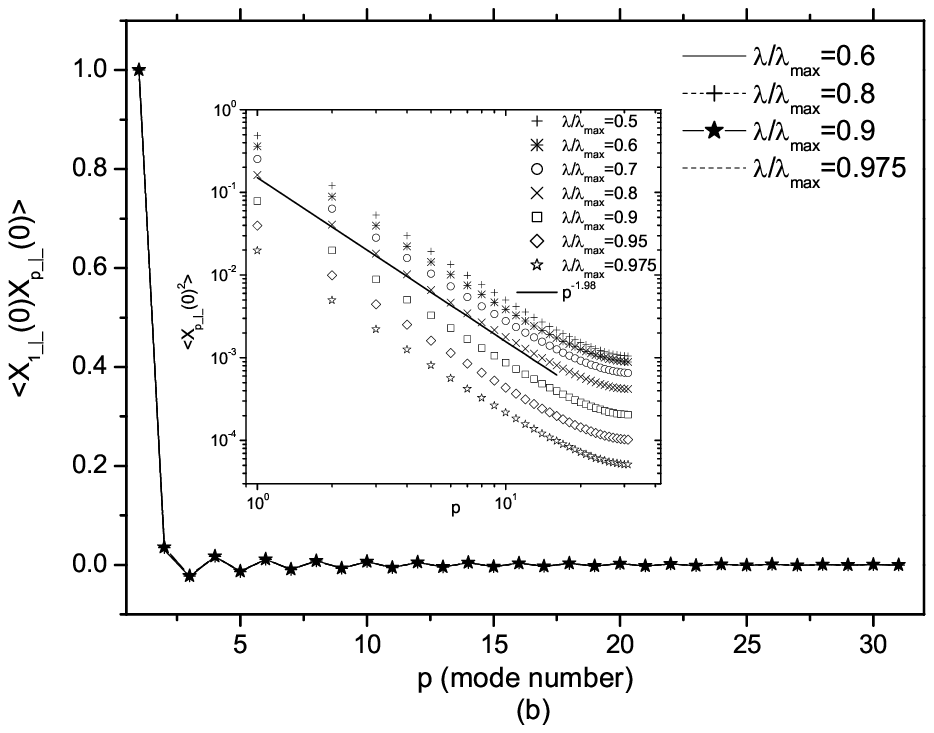}}
\caption{(a) $\langle X_{1\parallel}(0) X_{p\parallel}(0)\rangle$
against mode number $p$ for different relative extensions
$\lambda/\lambda_{max}$. In the inset we plot the $p$ dependence
of the function $\langle X_{p\parallel} (0)^2 \rangle$, we show
that the slope is almost equal to the predicted theoretical value
$\propto p^{-2}$ (b) The same as in (a) for the perpendicular
direction.}\label{fig:MCslope}
\end{figure}

An important question concerns the degree to which the stretching
affects the correlation between the first modes. In Figure
\ref{fig:MCslope} (a) and (b) we plot the autocorrelation $\langle
X_{p\parallel,\perp} (0)^2 \rangle$ and cross correlation $\langle
X_{1\parallel,\perp}(0) X_{m\parallel,\perp}(0) \rangle$ functions
as regards the mode number $m$ for different degrees of $\lambda /
\lambda_{max}$. We want to emphasize two features. First, one
should note that the power law of $\tau_{p\parallel,\perp}\propto
p^{-1.98}$ is observed for {\em all} extensions and in both
directions as this can be clearly seen from the insets of Fig.
\ref{fig:MCslope} (a) and (b). This behavior corresponds to our
theoretical predictions and reflects the fact that the chain
behaves like Gaussian at least as long as $0.5 < \lambda /
\lambda_{max} < 0.975$.

\begin{figure}[htb]
\centerline{\includegraphics[scale=0.85]{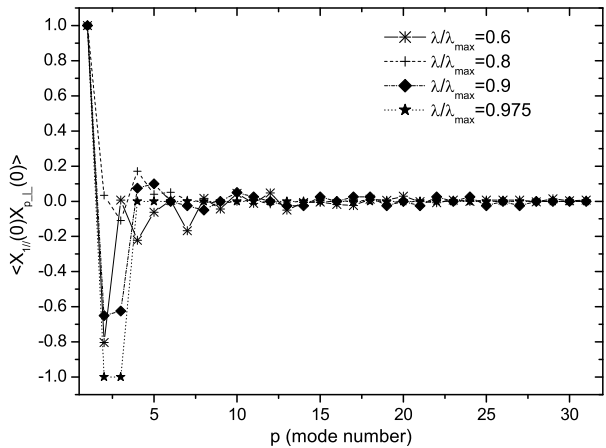}}
\caption{$\langle X_{1\parallel}(0) X_{p\perp}(0) \rangle$ as a
function of the mode number $p$ for different extensions $\lambda
/ \lambda_{max}$.} \label{fig:MCcorrZ1Xp}
\end{figure}

Second, we find that the modes of the same direction seem to be
orthogonal to each other for $1 < p < 31$, as in the linear
(Rouse) case. For the cross correlation terms between different
directions, $\langle X_{1\parallel}(0) X_{p\perp}(0)\rangle$, this
is not the case at least for the lower modes $1<p<10$ which
suggests that finite extensibility couples these modes in a
certain way, see Fig. \ref{fig:MCcorrZ1Xp}.

In conclusion, one may claim that this coupling effect, due to the
nonlinearity of the interactions, seems to be stronger between
modes of different directions and has apparently no effect on
modes in the same direction. Generally, only the first modes ($1 <
p < 10$) appear to couple since for modes with $p > 10$ (that is,
for short distance motions) the chain does not ``feel" the fixed
ends and the behavior is that of Rouse dynamics.

\subsection{Molecular Dynamics results}

Although real polymers do have mass, the inertia term is almost irrelevant
as compared with
the friction term, at least as long as the chain is not
extended,. For high extensions, however, due to the large
accelerations that each monomer is subject to, the inertia term is
comparable and even larger than the friction term and the whole
chain behaves as a string under tension in which oscillations are
no more overdamped.

In this section we check the relevance of the inertia term, which
should be important at the high stretching limit, and examine its
physical consequences. To this end we run Molecular Dynamics
simulations in which the mass of each monomer is accounted for in
the corresponding equations of motion. For the implementation of
the method we employ a successful numerical scheme developed and
proved earlier by Dimitrov et al.\cite{Dimitrov}. The model is
again that of a simple coarse-grained bead-spring chain,
originally proposed by Kremer and Grest \cite{Grest}, which has
been widely and very successfully used for MD simulations of
polymers in various contexts~\cite{KB,Kotel}. Effective monomers
along the chain are bound together by a combination of the FENE
potential, Eq. (\ref{fene}), where $K = 15$ and the maximum bond
extension is $l_{max} = 1.5\sigma$, and a Lennard-Jones potential,
responsible for the excluded volume interaction between the
monomers. $\sigma = 1$ is the range parameter of this purely
repulsive Lennard-Jones (LJ) potential which is truncated and
shifted to zero in its minimum and acts between any pairs of
monomers.
\begin{equation}\label{LJ}
 U_{LJ}({\bf r})=4\epsilon_{LJ}
\left[(\sigma/{\bf r})^{12}-(\sigma/{\bf r})^6+1\right], {\bf
r}\le {\bf r}_c=2^{1/6}\sigma
\end{equation}
The parameter $\epsilon_{LJ}$, characterizing the strength of this
potential, is chosen unity. Molecular Dynamics (MD) simulations
were performed using the standard Velocity-Verlet algorithm
\cite{Allen}, performing typically $1.5\times 10^9$ time steps
with an integration time step $\delta t = 0.01 t_0$ where the MD
time unit (t.~u.) $t_0 = (\sigma^2m / 48 \epsilon_{LJ} )^{1/2} =
1/ \sqrt{48}$, choosing the monomer mass $m_b=1$. The temperature
was held constant by means of a standard Langevin thermostat with
a friction constant $\zeta_0 = 0.5$

\subsubsection{MD Simulation results}

To make a comparison between Monte-Carlo and Molecular
Dynamics simulations, we study the same cases with the same
polymer chain of length $N = 32$. This time, $2^8$ MD steps in
each program running have been performed.

\begin{figure}[htb]
\centerline{\includegraphics[scale=0.85]{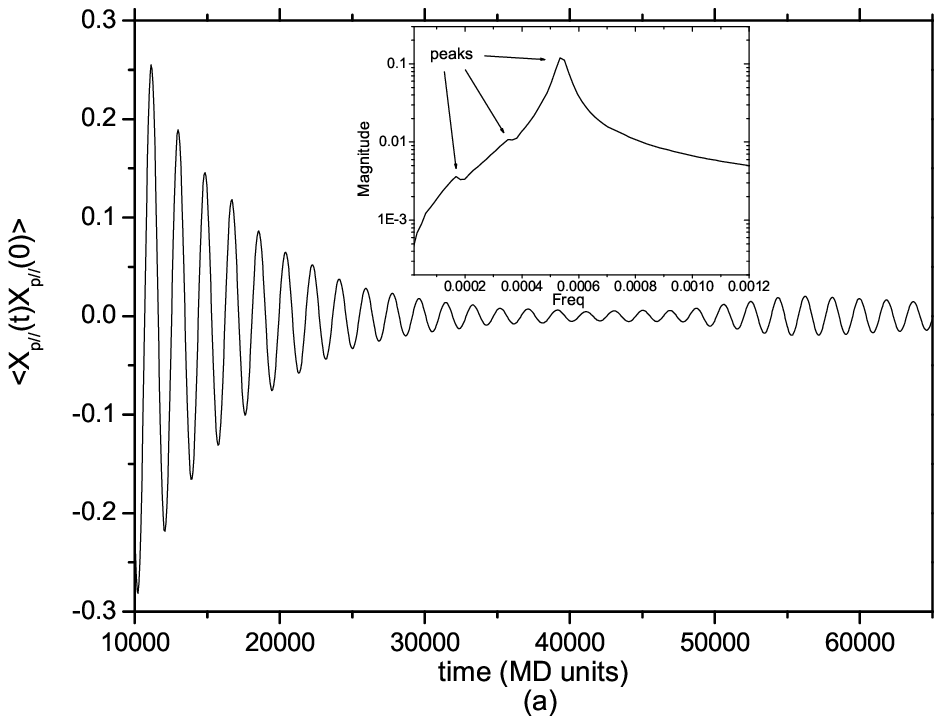}
\includegraphics[scale=0.85]{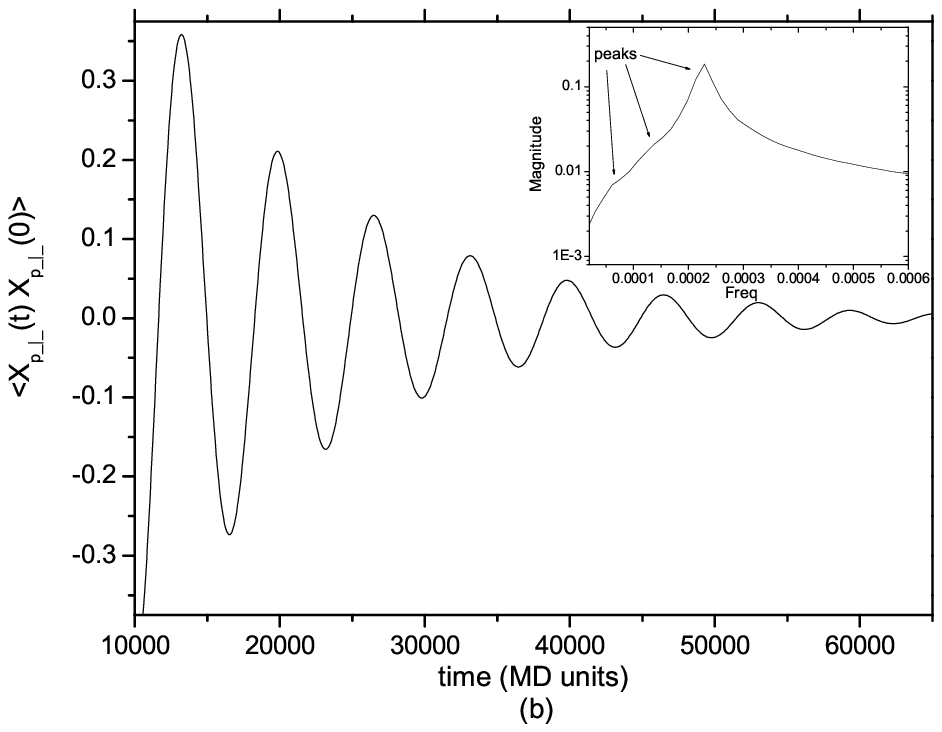}}
\caption{(a) $\langle X_{p\parallel}(t) X_{p\perp}(0)\rangle$
versus time (in MD time units) for $p = 3$. In the inset the
Fourier spectrum of the same function shows a sharp maximum,
corresponding to the high basic frequency.  Two smaller satellite
peaks correspond to lower frequencies, also present in the mode
dynamics. (b) The same as in (a) but in perpendicular direction.
The Fourier spectrum reveals the presence of a single well-defined
frequency in contrast of other two, less visible, smaller
peaks.}\label{fig:MDm3}
\end{figure}

Typical data for the time, equal-mode correlation functions
$C_{pp}(t) = \langle X_{p\parallel,\perp}(t)
X_{p\parallel,\perp}(0) \rangle$ are presented for a particular
value of $\lambda / \lambda_{max} = 0.83$ and for the mode $p =
3$. Figures \ref{fig:MDm3}a and \ref{fig:MDm3}b confirm the
importance of taking inertia into account, manifested by the
oscillations observed in the correlation functions. In Fig.
\ref{fig:MDm3}a, which corresponds to the mode in parallel
direction, one can recognize at least three frequencies in the
correlation function, exposed by smaller satellite peaks in the
Fourier spectrum in the inset. This broad-band spectrum does not
mean that only 3 frequencies are present in the correlation
function. One may rather claim that a continuum of frequencies of
decreasing amplitude appears instead,  reflecting the fact that
these frequencies are clearly discernable from their neighbor
values. Apparently, this effect manifests itself less visibly for
the perpendicular direction, at least for this value of $\lambda /
\lambda_{max}$ where only one peak can be well distinguished (see
Fig. \ref{fig:MDm3}b). This single frequency appearing in the
correlation function allows us to consider it as a linear {\em
underdamped} motion of the chain for this case.

\begin{figure}[htb]
\centerline{\includegraphics[scale=0.85]{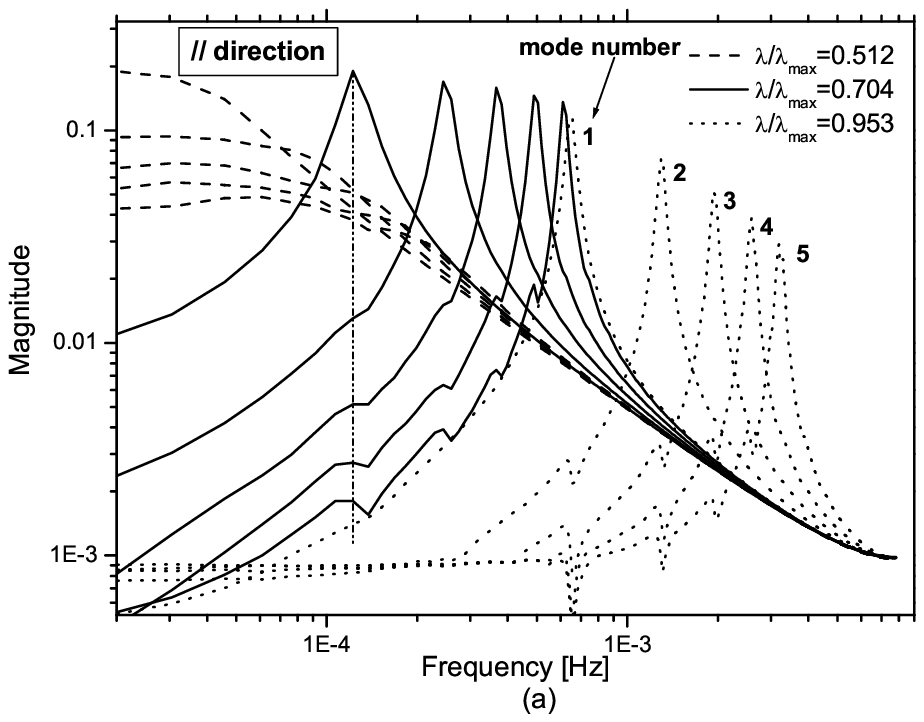}
\includegraphics[scale=0.85]{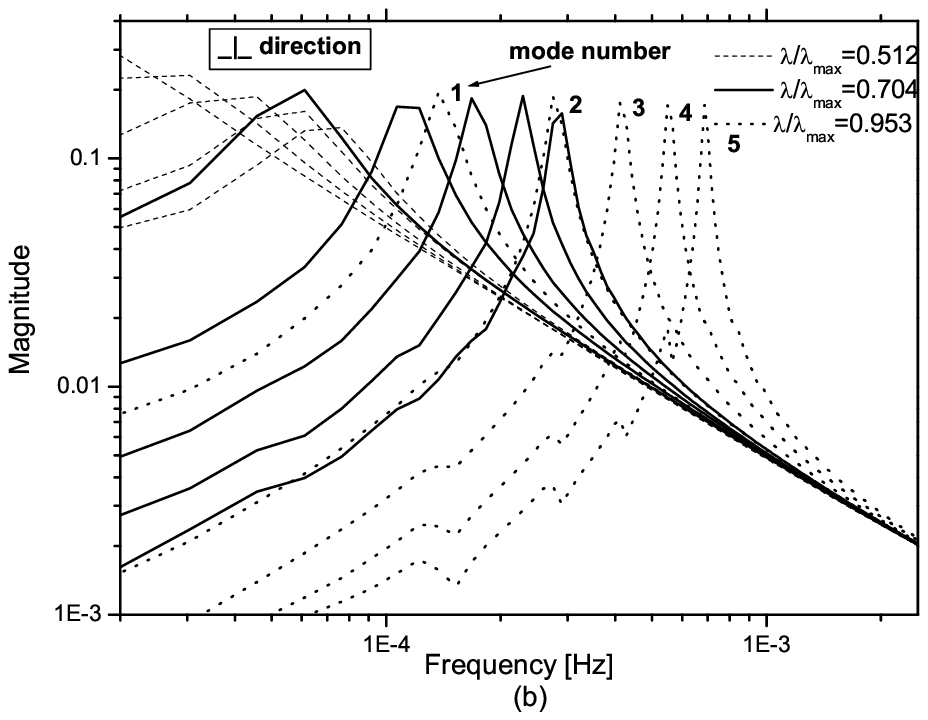}}
\caption{(a) Fourier spectra of correlation functions $\langle
X_{p\parallel}(t) X_{p\parallel}(0)\rangle$ for the first $5$
modes in parallel direction at three degrees of relative extension
$\lambda/\lambda_{max}$. The dash-dotted line is plotted as a
guide to the eye. (b) The same as in (a) for the perpendicular
direction.}\label{fig:MDfourier}
\end{figure}

At this stage we can ask for the physical meaning of these
satellite peaks shown in the above figure. To answer this
question, we plot together, in Figures \ref{fig:MDfourier}a and
\ref{fig:MDfourier}b the Fourier spectra (frequencies) of several
mode correlation functions at three different degrees of
extension. The first five modes in parallel direction are shown in
Fig. \ref{fig:MDfourier}a for $\lambda / \lambda_{max} = 0.512\;,
0.704\;, 0.953$. One can readily see that with growing stretching,
$\lambda / \lambda_{max}$, a central peak, corresponding to a
principal frequency of oscillation, is formed. However, this is
not the case for $\lambda/\lambda_{max} = 0.512$, where such a
peak is absent and the mode dynamics is spread almost uniformly
over frequencies lower that $10^{-4}$. In the other two cases
$\lambda / \lambda_{max} = 0.704\;, 0.953$ this spreading
disappears and narrower peaked functions come out, suggesting that
the mode motion is realized \emph{essentially} by means of a
single frequency. Now, we claim attention to a particular feature
of Figure \ref{fig:MDfourier}a. Take for example the case $\lambda
/ \lambda_{max} = 0.704$ for $p = 1$. There, by a vertical line we
demonstrate that the frequency corresponding to this mode ($p =
1$) coincides exactly with the satellite frequency peaks which
appear in the {\em higher} mode correlation functions. Looking
deeper, we can see that this is not prerogative of the first mode
and is observed also for $p > 1$. It takes place for larger
extensions too. Clearly, this shows that the satellite peaks are
exactly the main frequencies of higher mode correlation functions.
At the same time it indicates coupling of lower modes with higher
ones which is accompanied by exchange of energy between them.
Evidently, this coupling effect becomes more significant with
growing mode number and, most notably, at lager extensions (see
curves for $\lambda / \lambda_{max} = 0.953$).

Figure \ref{fig:MDfourier}b shows the same quantities for the
modes in perpendicular direction. In this case the situation is
similar whereby the satellite peaks indicate that coupling between
the modes is effectively weaker, and may be observed at the
largest extensions only.

\begin{figure}[htb]
\centerline{\includegraphics[scale=0.85]{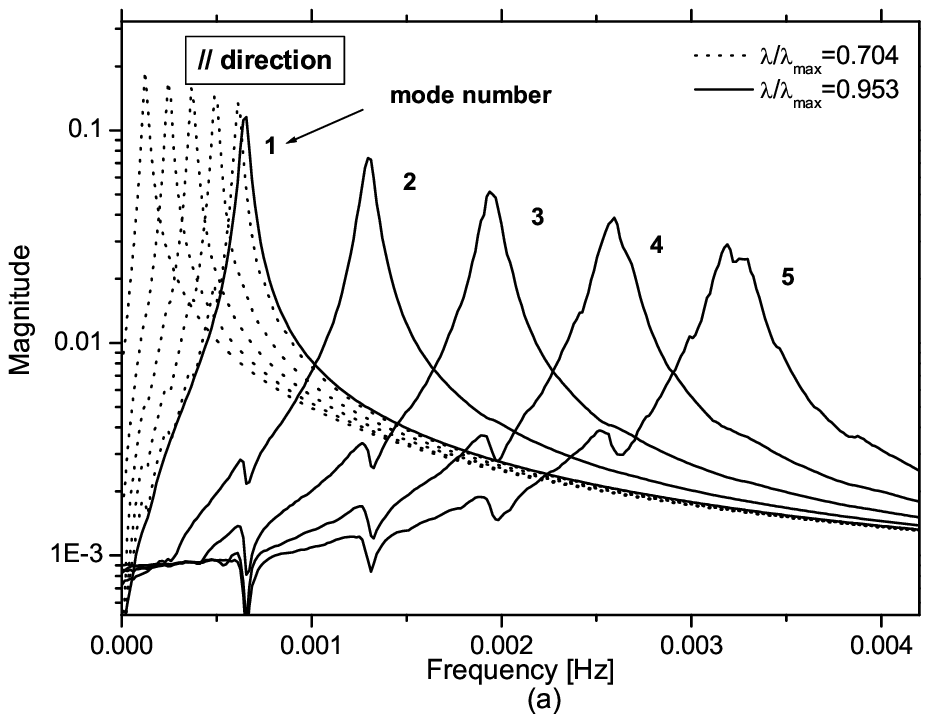}
\includegraphics[scale=0.85]{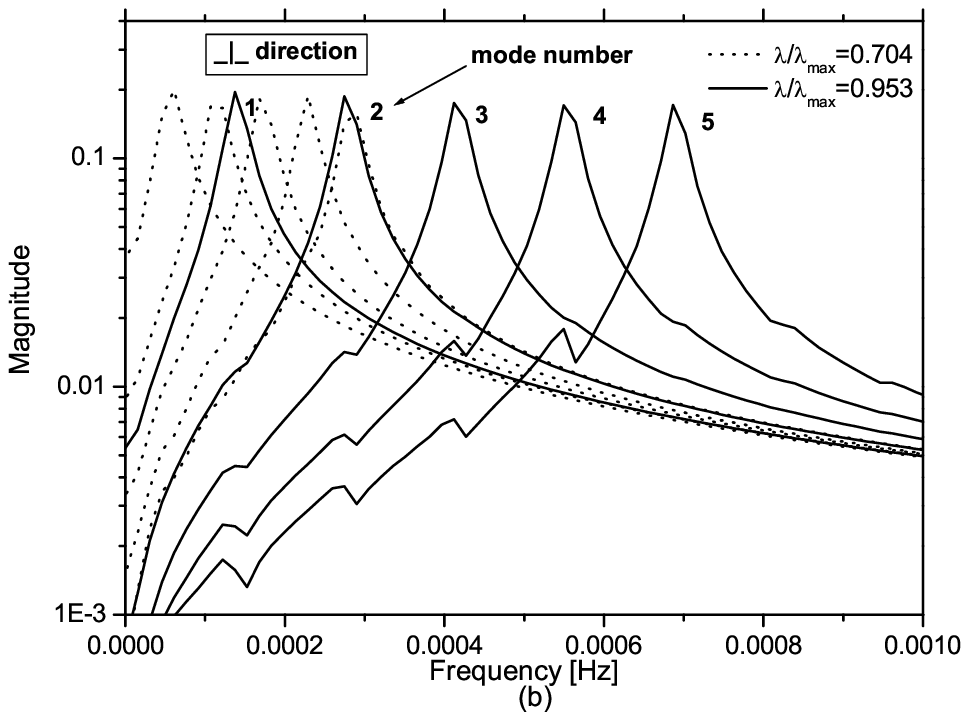}}
\caption{(a) Idem Figure \ref{fig:MDfourier}, where a linear scale
was used to show the broadening of the half-width of the main peak
in the Fourier spectra of correlation functions $\langle
X_{p\parallel}(t) X_{p\parallel}(0)\rangle$ for large values of
$\lambda/\lambda_{max}=0.704\; 0.953$.  (b) The same as in (a) but
this time the broadening if exists, does not appear to be
significant.}\label{fig:MDfourier-linear}
\end{figure}

More detailed information of the mode coupling effect can be
extracted if one plots, for higher extensions $\lambda
/\lambda_{max} = 0.704\;, 0.953$, the same functions as in Figure
\ref{fig:MDfourier}. This time, a normal scale in the frequency
axes (abscissas) is employed to facilitate visualization. In
Figure \ref{fig:MDfourier-linear}a, two significant effects can be
well distinguished. On the one hand, we can appreciate a slight
broadening of the half-width of the main peak with increasing
extension for the same mode. On the other hand, a similar
broadening occurs with increasing mode number, but this time at
constant extension. Also the height of the main peak decreases
with growing $p$. From a physical viewpoint, a broad-band spectra
centered in a given frequency is associated with a continuum of
frequencies (as mentioned above) around this single value.
Returning to our problem, this means that as soon as we increase
extension, more and more frequencies appear to make larger
contributions in the dynamics of single mode correlation
functions. A distinct situation accounts for the perpendicular
direction. Noticeably, this broadening is not seen in
Figure\ref{fig:MDfourier-linear}b, and the peak height remains
constant as one may verify from the similar graph depicted.

It is interesting to analyze the behavior of the principal
frequency $\omega$, corresponding to the peak in the Fourier
spectrum regarding the polymer chain extension.

\begin{figure}[htb]
\centerline{\includegraphics[scale=0.85]{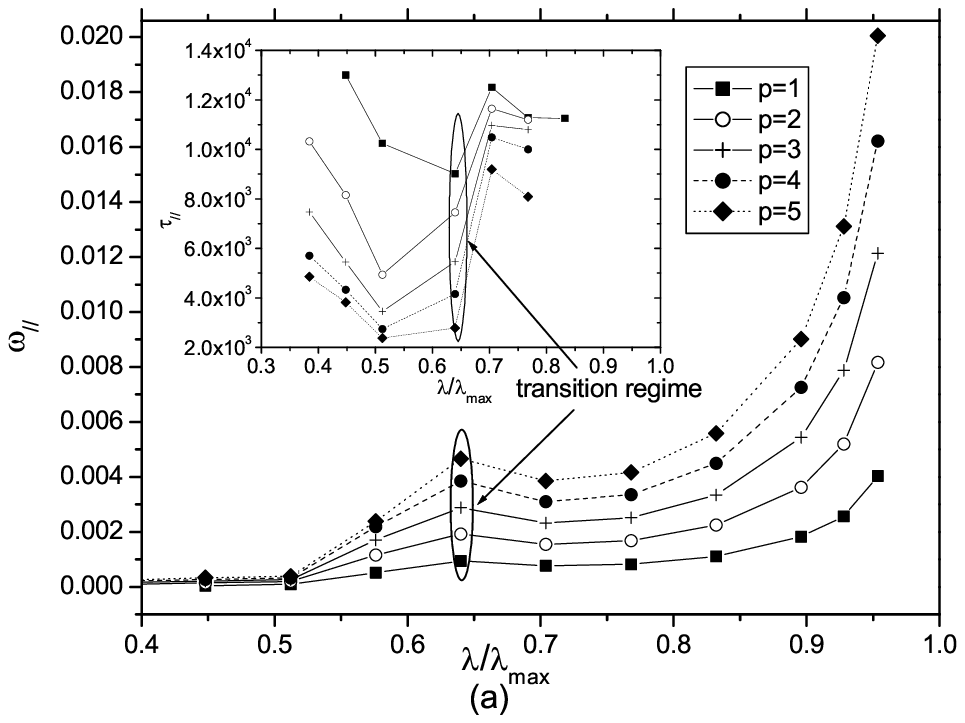}
\includegraphics[scale=0.85]{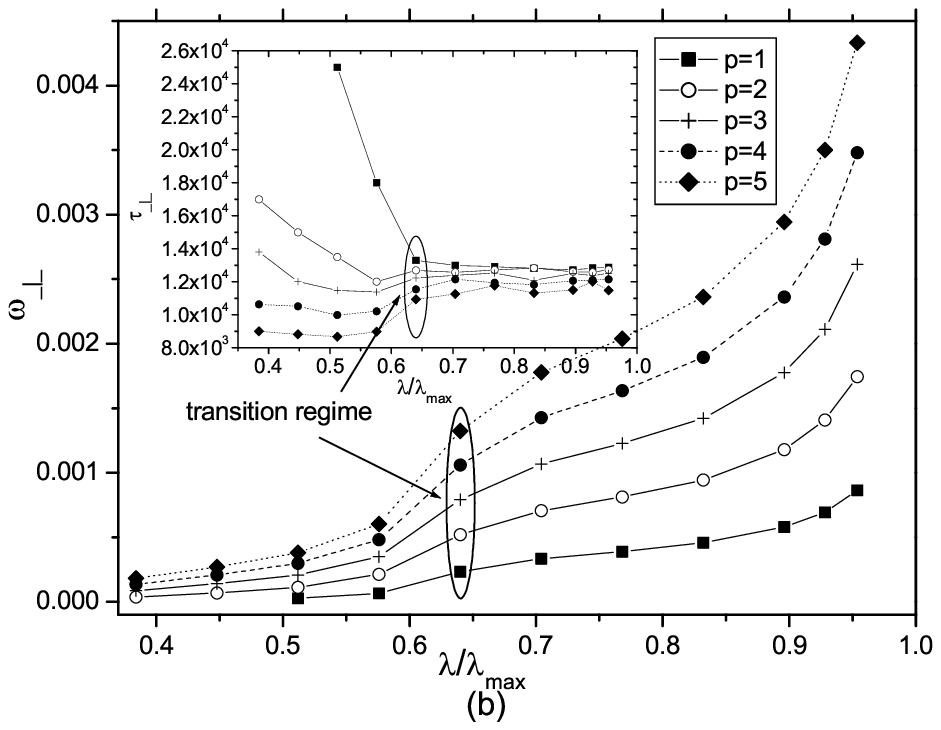}}
\caption{(a) Fourier spectrum of parallel mode correlation
functions, $\langle X_{p\parallel}(t) X_{q\parallel}(0)\rangle$
against chain extension. In the inset one shows the variation of
relaxation times $\tau_p$ for the first five modes with chain
elongation $\lambda / \lambda_{max}$. (b) The same as in (a) for
the perpendicular direction.}\label{fig:MDrelaxtime-omega}
\end{figure}

In Figure \ref{fig:MDrelaxtime-omega}a we demonstrate that the
characteristic frequency in parallel direction
$\omega_{\parallel}$ grows with increasing extension. At fixed
value of the elongation $\lambda / \lambda_{max}$,
$\omega_{\parallel}$ systematically increases with growing mode
number $p$. For a special value of $\lambda / \lambda_{max}\approx
0.64$, however, one observes a local maximum in
$\omega_{\parallel}$ which becomes increasingly pronounced as the
mode number $p$ increases. In the inset to Figure
\ref{fig:MDrelaxtime-omega}a, where we show the variation of the
relaxation times $\tau_p$ of the modes with changing elongation
$\lambda / \lambda_{max}$, the region around  $\lambda /
\lambda_{max} \approx 0.64$ marks the end of the steady decrease
of $\tau_p$ with growing degree of stretching. Thus, it appears
that at some threshold extension $\lambda / \lambda_{max} \approx
0.64$ the chain dynamics undergoes a qualitative transition from
an {\em overdamped} motion, determined essentially by friction, to
ballistic motion governed by inertial effects.

In the {\em overdamped} regime the MD results are qualitatively
consistent with the analytical predictions and MC data, indicating
a clear decay of the relaxation times with increased stretching.
Beyond some critical stretching, however, the  dynamics changes
and the polymer chain behaves increasingly like a string under
tension. Then  oscillations rather than random displacements of
the monomers become important and the relaxation times of the
different modes start to increase once again over a short range of
chain elongations. Upon further stretching these relaxation times
stay nearly constant. Eventually, in a regime of very strong
stretching for $\lambda/\lambda_{max} > 0.8$ the system becomes
strongly nonlinear in nature and it becomes difficult to determine
a well defined single ``relaxation time".

Figure \ref{fig:MDrelaxtime-omega}b shows similar plots for the
perpendicular direction. Rather than a maximum, the threshold
position at $\lambda / \lambda_{max} \approx 0.64$ marks the onset
of a stepper increase of the main frequency $\omega_{\perp}$ with
growing elongation whereby at $\lambda / \lambda_{max} \rightarrow
1$ this frequency appears to diverge $\omega_{\perp} \rightarrow
\infty$! The relaxation times $\tau_{\perp}$ of the various modes
in perpendicular direction behave differently  as compared to
their counterparts of Fig. \ref{fig:MDrelaxtime-omega}a. While in
the {\em overdamped} regime at $\lambda / \lambda_{max} < 0.64$
they decay steadily with growing elongation, in the inertial
regime here after the threshold they merge to a nearly single
value which remains largely unchanged for $\lambda / \lambda_{max}
> 0.64$. Remarkably, since the oscillations of the string in perpendicular
direction persist up to very large extensions, $\lambda /
\lambda_{max} \approx 1$, a well defined relaxation time can be
found.

\subsubsection{Approximate analytical model}

It is possible to explain the observed behavior of the principal
frequencies in the Fourier spectra of the mode-mode correlation
functions as well as that of the relaxation times if one considers
the linearized equations of motion (\ref{eq:firstmoment}). The
intention here is to capture the essential features of the main
frequencies $\omega_{\parallel,\perp}$ dependence on mode number
$p$ and understand the insensitivity of the relaxation times with
regard to chain elongation.

Our model of a chain pulled by force $f$ and fixed at the origin
allow us to think of the chain as being fixed in space with
elongation $\bar{R}_{\parallel}(N)$. One can then write an
approximate equation of motion for ${\bf R}(s,t)$ using the
information provided by the GSC method of section \ref{sec:GSC}.

First, we will decompose the position vector as ${\bf R}(s,t) =
\widehat{{\bf R}}(s,t)+\langle {\bf R}(s,t) \rangle$,  where
$\widehat{{\bf R}}(s,t)$ is a new vector measuring the deviations
of the position vector from its average position $\langle {\bf
R}(s,t) \rangle$. Then, making use of previously defined
quantities and expression (1), we can propose the following
linearized equations
\begin{equation}\label{eq:firstmoment1}
m_b \left\langle\frac{\partial^2 R_{j}(s,t)}{\partial t^2}
\right\rangle + \xi_0 \left\langle\frac{\partial R_{j} (s,t)}
{\partial t} \right \rangle - K_j \left \langle \frac{\partial^2
R_{j}(s,t)} {\partial s^2} \right \rangle - f \delta_{sN}
\delta_{jz} = 0 \hspace{0.5in} j=x,\;y,\;z
\end{equation}
where we have replaced the exact force by a linearized
(approximate) elastic force with elastic constants $K_j =
K_{\parallel}, K_{\perp}$, obtained in section
\ref{sec:steady-state}

\begin{equation}
\left[ \left \langle \frac {\delta V[{\bf R}(s,t) - {\bf
R}(s-1,t)]} {\delta R_{\parallel, \perp}(s,t)} \right \rangle +
\left \langle \frac {\delta V[{\bf R}(s,t) - {\bf R}(s+1,t)]}
{\delta R_{\parallel, \perp}(s,t)} \right \rangle \right] \approx
K_{\parallel, \perp} \left \langle \frac {\partial^2 R_{\parallel,
\perp}(s,t)} {\partial s^2} \right \rangle
\end{equation}
Separating the problem in the parallel and perpendicular direction
as before and using the previously defined decomposition of vector
${\bf R}(s,t)$, since equations (\ref{eq:firstmoment1}) are
linear, we finally arrive at
\begin{equation}\label{eq:linearfirstmpar-perp}
m_b \left \langle \frac {\partial^2 \widehat{R} _{\parallel,
\perp}(s,t)} {\partial t^2} \right \rangle+ \xi_0 \left \langle
\frac {\partial \widehat {R}_{\parallel, \perp} (s,t)} {\partial
t} \right \rangle- K_{\parallel, \perp} \left \langle \frac
{\partial^2 \widehat {R}_{\parallel,\perp}(s,t)} {\partial s^2}
\right \rangle = 0
\end{equation}
Introducing the Fourier expansion for $\langle \widehat
{R}_{\parallel, \perp}(s,t) \rangle = \sum_{p} \langle \widehat
{X}_{p\parallel, \perp} (t) \rangle \sin \left ( \frac{p \pi s}
{N} \right)$ and inserting it in (\ref{eq:linearfirstmpar-perp}),
we obtain  after multiplying it by $\widehat {X}_{p\parallel,
\perp}(0)$
\begin{equation}\label{eq:summodes} \sum_p \left(m_b
\ddot {\widehat {C}}_{p\parallel, \perp} (t) + \xi_0 \dot
{\widehat {C}}_{p\parallel, \perp} (t) + K_{\parallel, \perp}
\left ( \frac{p \pi}{N} \right)^2 \widehat {C}_{p\parallel,
\perp}(t) \right ) \sin \left ( \frac {p \pi n} {N}\right) = 0
\end{equation}
where $\dot {\widehat {C}}_{p\parallel, \perp} (t) \equiv d
\widehat {C}_{p
\parallel, \perp} (t) / dt$ and $\widehat {C}_{p \parallel, \perp} (t) = \langle
\widehat {{X}}_{p \parallel, \perp} (t) \widehat {{X}}_{p
\parallel, \perp} (0) \rangle$. To satisfy (\ref{eq:summodes}), we
require that
\begin{equation}\label{eq:summodesres}
m_b  \ddot {\widehat {C}}_{p \parallel, \perp}+ \xi_0 \dot {
\widehat{C}}_{p
\parallel, \perp} + K_{\parallel, \perp} \left ( \frac{p \pi} {N} \right)^2
\widehat {{C}}_{p \parallel, \perp} = 0
\end{equation}
Assuming that $\widehat {C}_{p \parallel,\perp} (t) = C_{p
\parallel, \perp}^0 \exp {(-\gamma_{p \parallel, \perp}t)}$, we
finally obtain the eigenvalues of the linearized problem
\begin{equation}\label{eq:gamma}
\gamma_{p \parallel, \perp} = \frac { \xi_0} {2 m_b}\pm \sqrt {
\frac {\xi_0^2} {4 m_b^2} - \frac {K_{\parallel, \perp}} {m_b}
\left ( \frac {p \pi} {N} \right)^2}
\end{equation}
Clearly, these eigenvalues are real, showing a purely exponential
decay, or complex, indicating damped oscillations, depending on
the value of the radical in (\ref{eq:gamma}). Hence, these cases
must be treated separately.

\begin{itemize}
     \item \emph{Overdamped case}
\end{itemize}

In this case, the radical is real, provided
\begin{equation}\label{eq:overdamped}
\left ( \frac {\xi_0} {2 m_b} \right)^2 > \frac {K_{\parallel,
\perp}} {m_b} \left ( \frac {p \pi} {N} \right)^2
\end{equation}
and it is possible to identify $\gamma_{p \parallel, \perp}$ with
the inverse of the relaxation time $\gamma_{p \parallel, \perp} =
1 / \tau_{p \parallel, \perp}$.

Returning to (\ref{eq:gamma}) and retaining only the first term in
the Taylor expansion of the radical, $\sqrt {1-x^2} \approx 1
-\frac {1} {2} x^2$, we find two values of $\gamma_{p\parallel,
\perp}$ or $\tau_{p\parallel, \perp}$:
\begin{equation}\label{eq:omegaover}
\frac {1} {\tau_{p \parallel, \perp}^{(1)} } \approx \frac
{K_{\parallel, \perp}} {\xi_0} \left ( \frac {p \pi} {N} \right
)^2 \hspace{0.2in} \text{and} \hspace {0.2in} \frac {1} {\tau_{p
\parallel, \perp}^{(2)}} \approx \frac {\xi_0} {m_b} - \frac
{K_{\parallel, \perp}} {\xi_0} \left ( \frac {p \pi} {N} \right
)^2
\end{equation}
since eq. (\ref{eq:summodesres}) is of second order in time.
Evidently, to recover the theoretical result of section
\ref{sec:GSC} for the relaxation time $\tau_{p\parallel, \perp}$,
one must neglect the inertia term (inserting, for example, $m_b =
0$) in (\ref{eq:firstmoment1}) whereby only $\tau_{p \parallel,
\perp}^{(1)}$ survives.

\begin{itemize}
     \item \emph{Underdamped case}
\end{itemize}

In this case the radical is negative,
\begin{equation}\label{eq:underdamped}
\left ( \frac {\xi_0} {2 m_b} \right )^2 < \frac {K_{\parallel,
\perp}} {m_b} \left ( \frac {p \pi} {N} \right)^2
\end{equation}

Now, $\gamma_{p \parallel, \perp}$'s  are complex quantities
meaning that $ \widehat{C}_{p \parallel, \perp} (t)$ corresponds
to damped oscillatory motion. Clearly, the  real part of
$\gamma_{p\parallel, \perp}$ describes the degree of damping and
the imaginary one gives the oscillation frequency which can be
called $\omega_{p \parallel, \perp}$. If we still consider the
real part of $\gamma_{p \parallel, \perp}$ as the inverse of a new
relaxation time $\tau_{p \parallel, \perp }^{u}$ for the {\em
underdamped} case, we find that $\gamma_{p \parallel, \perp}$ is
equal to
\begin{figure}[htb]
\centerline{\includegraphics[scale=0.8]{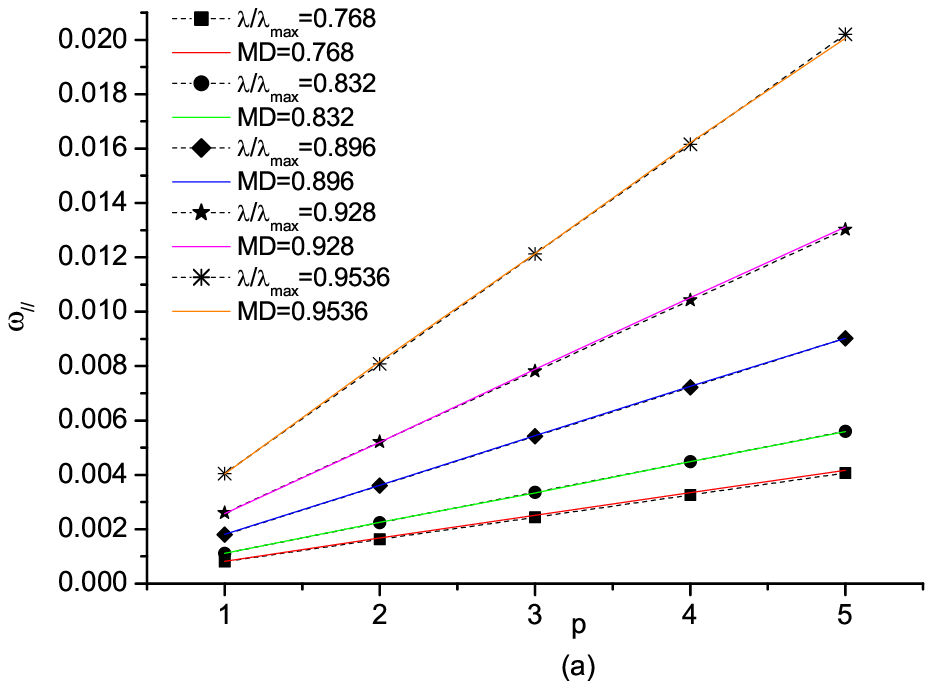}
\includegraphics[scale=0.8]{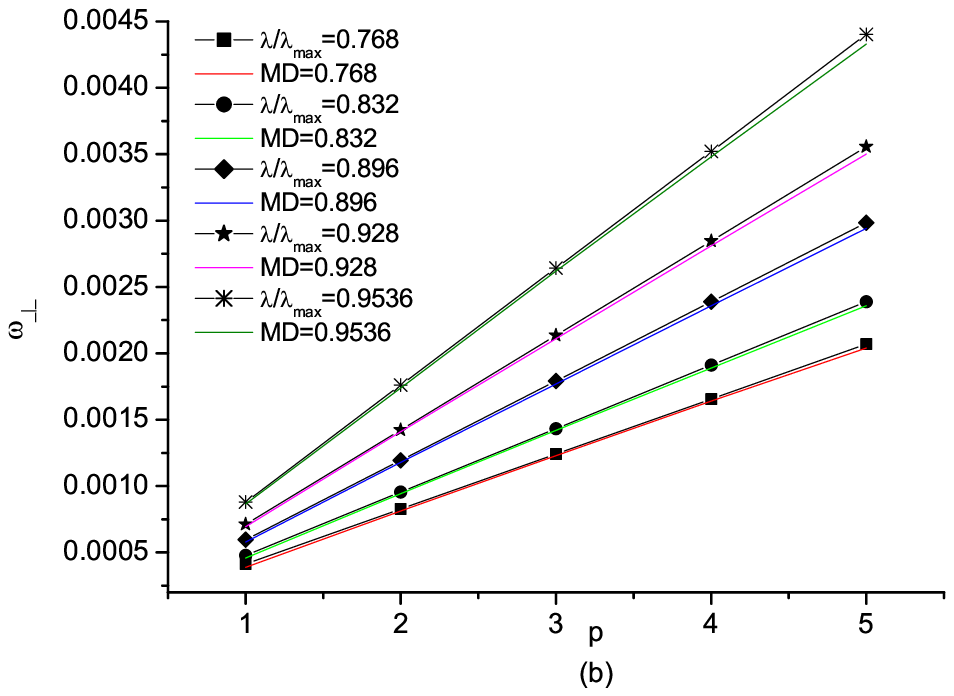}} \caption{ (a) Main
frequency of the parallel mode correlation function $\omega_{p
\parallel}$ as a function of mode number $p$ for different
extensions in the \emph{underdamped} regime. In the legend
$\lambda / \lambda_{max}$ means analytical approximation, MD ,
Molecular dynamics simulations. (b) The same as in (a) in the
perpendicular direction.}\label{fig:freqcies}
\end{figure}

\begin{equation}\label{eq:omegaunder}
\gamma_{p \parallel, \perp} = \frac {1} {\tau_{p \parallel,
\perp}^{u}} \pm i \omega_{p \parallel, \perp} = \frac {\xi_0} {2
m_b} \pm i \left ( \frac {K_{\parallel, \perp}} {m_b} \right
)^{1/2} \left ( \frac{p \pi} {N} \right) \sqrt {1- \left ( \frac
{\xi_0} {2 m_b} \right)^2 \left ( \frac{N} {p \pi} \right )^2
\frac {m_b} {K_{\parallel, \perp}}}
\end{equation}
where $i = \sqrt{-1}$ is the imaginary unit. This can be
approximated for large $p$ or large extensions ($K_{\parallel,
\perp} \gg 1$) by
\begin{equation}\label{eq:omegaunderapprox}
\frac {1} {\tau_{p \parallel, \perp}^{u}} = \frac {\xi_0} {2 m_b}
\hspace{0.2in} \text{and} \hspace {0.2in} \omega_{p \parallel,
\perp} \approx \pm \left ( \frac {K_{\parallel, \perp}} {m_b}
\right )^{1/2} \left ( \frac {p \pi} {N} \right )
\end{equation}
where a Taylor expansion of the radical up to first order was
carried out.

The last equation reveals some remarkable features. First of all,
we can see a {\em linear} dependence on $p$ of the main
frequencies in the mode correlation functions $\omega_{p
\parallel, \perp}$. Secondly, an increasing of the magnitude of
these frequencies upon increased stretching is also demonstrated.
Figures \ref{fig:freqcies}a and b support these observations
showing the reliability of the approximations. We plot $\omega_{p
\parallel}$ \ref{fig:freqcies}a and $\omega_{p \perp}$
\ref{fig:freqcies}b against mode number $p$ for different
extensions. Note that these findings apply only to the
\emph{underdamped} regime, $\lambda / \lambda_{max} > 0.64$.

Moreover, a nearly constant relaxation time, $\tau_{p \parallel,
\perp}^{u}$, which doesn't depend on mode number or extension,
comes out from equation (\ref{eq:omegaunderapprox}) too. Upon
qualitative comparison between Figures
\ref{fig:MDrelaxtime-omega}a and b (inset plots), we can conclude
that for the perpendicular direction \ref{fig:MDrelaxtime-omega}b,
where a single relaxation time can be well defined even at very
high extensions $\lambda / \lambda_{max} \approx 0.95$ the
agreement is good. Instead, for the parallel direction
\ref{fig:MDrelaxtime-omega}a, a constant value of the relaxation
time agrees with simulation results up to the extension where a
single relaxation time can be well defined.

\section{Conclusions}

In the present paper we have studied the dynamics of a
self-avoiding polymer chain at different degrees of stretching.
Analytical work as well as computer simulations provide a rather
consistent picture of the qualitative changes which the polymer
dynamics undergoes upon gradual increase of the degree of
stretching. In general, one observes two regimes of chain
dynamics, depending on the degree of chain extension. In the first
one, that of the friction dominated \emph{overdamped} motion of
the monomers, both analytic predictions as well as Monte Carlo
results suggest a consistent picture of relaxation time decrease
with growing stretching of the chain up to a threshold value
$\lambda/\lambda_{max} \approx 0.64$ whereby the relaxation time
parallel to stretching, $\tau_{||}$ is always considerably smaller
(about one half) of that in direction perpendicular to stretching,
$\tau_{\perp}$.  For $\tau_{||}$ the agreement between analytic
results from the GSC approximation and MC data is perfect on a
quantitative level whereas for $\tau_{\perp}$ it is at most
qualitative. As expected, the MC results for the relaxation time
versus mode index relationship yield $\tau_p \propto p^{-2}$.

A {\em transition} regime at $\lambda / \lambda_{max} \approx
0.64$ where friction and inertia terms are of the same order of
magnitude separates the \emph{overdamped} regime from an inertial
regime at higher degrees of stretching. This latter regime is
strongly nonlinear in nature and a faithful description of the
polymer dynamics can be produced by means of a MD simulation as
well as by an approximate linearized analytical model.

Among the most salient features of polymer dynamics in this second
regime of strong stretching, we find that normal modes are coupled
to each other and can interchange energy, as expected in a system
with strongly anharmonic interactions. These anharmonic
contributions to the bond (FENE)-potential come into play at
strong stretching only whereas in a weakly extended chain the
monomers make the bonds oscillate around the equilibrium bond
length by means of nearly harmonic forces and the modes are
largely independent.

The Fourier analysis reveals the existence of principal
characteristic frequencies of oscillatory motion for each normal
mode. Along with this feature, a broad-band spectrum suggest the
existence of a continuum of frequencies around the single peaks,
which apparently make a non-vanishing contribution to the total
motion. These principal characteristic frequencies have been shown
by an approximate linear analytical model to scale {\em linearly}
with mode index $p$. The frequencies also grow upon extension, as
expected. A slight broadening of the half-width of the main
frequency peak with increasing extension as well as with
increasing mode number at constant extension, is revealed for the
parallel direction. This means that as soon as we increase
extension, more and more frequencies appear to contribute in the
polymer dynamics, reflected by the single mode correlation
functions. This broadening is not evidenced, at least up to
$\lambda / \lambda_{max} = 0.95$, for the perpendicular direction.

The relaxation times of the modes depend significantly on the
considered direction. While in the parallel direction the
relaxation times increase strongly in the transition region but
then the possibility to define a single relaxation time for too
large extensions $\lambda / \lambda_{max} > 0.8$ becomes
questionable, in the perpendicular direction the relaxation times
remain more or less constants up to $\lambda / \lambda_{max}
\approx 0.95$. It is remarkable that these results are also
confirmed qualitatively by our linearized approach. Of course, one
must keep in mind that  such large extensions of a polymer chain
might be of academic interest since bond rupture make take place
if one approaches the tensile strength of the macromolecule.

\section{Acknowledgements}

M. F. acknowledges the support and great hospitality of the Max
Planck Institute for Polymer Research in Mainz, during his
postdoctoral research position and also to the DFG FOR 597 for
financial support. He also is indebted to Universidad Nacional del
Sur and CONICET (Argentina) which gave him the possibility to make
his postdoctoral studies. D.D. appreciates support from the Max
Planck Institute of Polymer Research via MPG fellowship. A. M. and
V.R. acknowledge support from the Deutsche Forschungsgemeinschaft
(DFG), grant No. SFB 625/B4.

\newpage
\appendix
\section{}\label{ap:2}
In calculating terms like $\left \langle \frac{\delta V[{\bf
R}(s,t) - {\bf R}(s-1,t)]}{\delta R_{\parallel}(s,t)}\right
\rangle$ one needs to tackle the term $\langle \delta [{\bf
R}(s,t) - {\bf R}(s-1,t)-r] \rangle$ - see Section
\ref{sec:Gauss-preav}. To this end we make the following
approximation
\begin{equation}
 \langle \delta[{\bf R}(s,t) - {\bf R}(s-1,t)-r]\rangle \approx
\delta [ \langle {\bf R}(s,t) - {\bf R}(s-1,t)-r \rangle ].
\end{equation}
Taking into account that $V(\vec{r}) = - \frac{1}{2} k_F b_0^2
\ln(1 - \frac{\vec{r}^2}{b_0^2})$ and $\vec{r}^2 = r_{\parallel}^2
+ r_{\perp}^2$, one can set

\begin{eqnarray*}
 \int d^3 r \langle \delta [ {\bf R}(s,t) - {\bf R}(s-1,t)-r] \rangle
\frac{\partial V(\vec{r})}{\partial r_{\parallel}} \approx k_F
\frac {\bar{R}_{\parallel} (s,t)-\bar{R}_{\parallel}(s-1,t)} {1 -
\left [ \frac {\bar{R}_{\parallel}(s, t) -
\bar{R}_{\parallel}(s-1,t)} {b_0}\right]^2}
\end{eqnarray*}

which is the final expression that must be inserted in the
equation of motion.

The calculation of $A_{\parallel}(s,n,t) = C_{\parallel}(s,n,t) -
\bar{R}_{\parallel}(s,t) \bar{R}_{\parallel}(n,t)$, proceeds as
follows:

From equation (\ref{eq:eqtcorrparant}) we have to calculate terms
like $\left \langle R_{\parallel}(n,t) \frac{\delta V[ {\bf
R}(s,t) - {\bf R}(s-1,t)]} {\delta R_{\parallel}(s,t)} \right
\rangle$. Using the $\delta$-function, one can write

\begin{eqnarray}\label{eq:ap_correl}
 \left \langle R_{\parallel}(n,t) \frac{\delta V[{\bf R}(s,t) - {\bf R}(s-1,t)]}
{\delta R_{\parallel}(s,t)} \right \rangle = \left \langle
R_{\parallel}(n,t) \int d^3 r \langle \delta [ {\bf R}(s,t) - {\bf
R}(s-1,t)-r] \rangle \frac{\partial V(\vec{r})}{\partial
r_{\parallel}} \right \rangle
\end{eqnarray}

Applying Wick's theorem \cite{Wick} we have

\begin{small}
\begin{eqnarray*}
 \left \langle R_{\parallel}(n,t) \frac{\delta V[{\bf R}(s,t) - {\bf R}(s-1,t)]}
{\delta R_{\parallel}(s,t)} \right \rangle = \langle
R_{\parallel}(n,t) \rangle \int d^3 r \langle \delta [ {\bf
R}(s,t) - {\bf R}(s-1,t)-r] \rangle \frac{\partial V(\vec{r})}
{\partial r_{\parallel}} + (\langle
R_{\parallel}(n,t) R_{\parallel}(s,t) \rangle \\
- \langle R_{\parallel}(n,t) \rangle \langle
R_{\parallel}(s,t)\rangle) \left \langle \frac{\delta}{\delta
R_{\parallel}(s,t)} \int d^3 r \delta [{\bf R}(s,t) - {\bf
R}(s-1,t)-r] \frac{\partial V(\vec{r})} {\partial r_{\parallel}}
\right
\rangle + (\langle R_{\parallel}(n,t) R_{\parallel}(s-1,t) \rangle \\
- \langle R_{\parallel}(n,t) \rangle \langle R_{\parallel}(s-1,t)
\rangle)\left \langle \frac{\delta}{\delta R_{\parallel}(s - 1,t)}
\int d^3 r \delta [ {\bf R}(s,t) - {\bf R}(s-1,t)-r]
\frac{\partial V(\vec{r})} {\partial
r_{\parallel}}   \right \rangle\\
= \bar{R}_{\parallel}(n,t) \int d^3 r \langle \delta [ {\bf
R}(s,t) - {\bf R}(s-1,t)-r] \rangle \frac{\partial
V(\vec{r})}{\partial r_{\parallel}} +
[A_{\parallel}(n,s,t)-A_{\parallel}(n,s-1,t)] \int d^3 r \langle
\delta [ {\bf R}(s,t) - {\bf R}(s-1,t)-r] \rangle \frac{\partial^2
V(\vec{r})} {\partial r_{\parallel}^2}
\end{eqnarray*}
\end{small}
where the rhs was obtained after integrating by parts. For the
term $\left \langle  R_{\perp}(n,t) \frac{\delta V[{\bf
R}(s,t)-{\bf R}(s-1,t)]}{\delta R_{\parallel}(s,t)}\right
\rangle$, one has to replace the parallel variables by the
perpendicular counterpart, taking into account that
$\bar{R}_{\perp}(s) \equiv 0$.

Concerning the meaning of $K_{\parallel, \perp}(s,s \pm 1)$, we
define an effective elastic constant:

\begin{equation}\label{Constant_K}
K_{\parallel,\perp}(s,s \pm 1) = \int d^3 r \langle \delta ({\bf
R}(s,t) - {\bf R}(s \pm 1,t) - r) \rangle \frac{\partial ^2
V(r)}{\partial r_{\parallel,\perp}^2}
\end{equation}
Applying the same approximation as before, and after taking the
second derivatives of the potential $V(r)$ with respect to
perpendicular and parallel directions, we have for the
perpendicular effective constant
\begin{equation}\label{eq:Kperp}
K_{\perp}(s,s \pm 1) = k_F \frac {1} {1-\left [
\frac{\bar{R}_{\parallel}(s,t) - \bar{R}_{\parallel}(s \pm
1,t)}{b_0} \right ]^2}
\end{equation}
and for the parallel one
\begin{equation}\label{eq:Kpar}
K_{\parallel}(s,s \pm 1) = k_F \frac {1 + \left [ \frac
{\bar{R}_{\parallel} (s,t) - \bar{R}_{\parallel} (s \pm 1,t)}{b_0}
\right ]^2} {\left (1 - \left [\frac {\bar{R}_{\parallel}(s,t) -
\bar{R}_{\parallel}(s \pm 1,t)}{b_0} \right ]^2 \right )^2}
\end{equation}

The expression for the term $\langle R_{\perp}(n,t) h_{\perp}(s,t)
\rangle$  can also be calculated by making use of the generalized
Wick's theorem \cite{Wick}:

\begin{equation}
\langle R_{\parallel,\perp}(n,t) h_{\parallel,\perp}(s,t) \rangle
= 2 \xi_0 k_B T \left \langle \frac{ \delta
R_{\parallel,\perp}(n,t)} {\delta h_{\parallel,\perp}(s,t)} \right
\rangle
\end{equation}
As can be seen in \cite{Wick}, the term $\left \langle \frac{
\delta R_{\parallel,\perp}(n,t) }{\delta h_{\parallel,\perp}(s,t)}
\right \rangle = \frac{1}{2 \xi_0} \delta_{ns}$ so that the
previous term gives

\begin{equation}
 \langle R_{\parallel,\perp}(n,t) h_{\parallel,\perp}(s,t) \rangle = \delta_{ns} k_B T
\end{equation}

\end{document}